\documentclass[preprint,nofootinbib]{revtex4-1}

\usepackage{amsmath}
\usepackage{verbatim}
\usepackage{mathrsfs}
\usepackage{amssymb}
\usepackage{amsfonts}

\usepackage{graphicx}
\usepackage[below]{placeins}

\usepackage{natbib}
\usepackage[english]{babel}

\begin{document}

\title{Revisiting Higgs inflation in the context of collapse theories}
\author{Saul Rodriguez}
\email{saul.rodriguez@nucleares.unam.mx}
\author{Daniel Sudarsky}
\email{sudarsky@nucleares.unam.mx}
\affiliation{National Autonomous University of Mexico}

\begin{abstract}
	In this work we consider the Higgs inflation scenario,  but in contrast with past  works,  the  present   analysis  is   done  in the context   of  a  spontaneous  collapse theory   for the  quantum state of  the  inflaton  field.  In particular, we  will rely on a  previously studied  adaptation of  the Continuous Spontaneous Localization  model for the  treatment of  inflationary  cosmology. We will show that  with the  introduction of  the dynamical collapse hypothesis, some of the most serious   problems of the  Higgs inflation  proposal  can  be  resolved  in a natural  way.
\end{abstract}

\maketitle

\section{Introduction}

The inflationary paradigm has become an integral part of modern  cosmology,  in great  part due  to its empirically successful account of the primordial inhomogeneities that represent the origin of the cosmological structure. The most recent observational data of the CMB,   as  extracted  from the  PLANCK satellite  observations, is consistent with the theoretical predictions made by   many inflationary models \cite{Hinshaw:2012aka,Ade:2014xja}, including some  of  the single scalar field models \cite{Martin:2013nzq}.  There is  an  outstanding problem regarding the  still unobserved   primordial tensor perturbations and   we point the reader to \cite{Leon:2013pya}  for  a   re-assessment of the relevant  expectations   based  on ideas  related to the ones underlying the present work. 

According to  the inflationary paradigm, the  explanation  for  the  generation of primordial inhomogeneities  that constitute the seeds of cosmic  structure is the following: the perturbations start as quantum fluctuations  associated  with the vacuum state  of the inflaton field,  as the universe goes through an era of accelerated expansion. The physical wavelength associated with the perturbations is stretched out by the expansion, eventually reaching cosmological scales. At this point, quantum fluctuations  are  treated  as  describing the averages over an ensemble of inhomogeneous universes of their analogue classical quantities: classical density perturbations \cite{Mukhanov:1990me}. That  last  step, however  does not seem to have a  clear justification,  as it has  been  noted  in \cite{Perez:2005gh}   and  pointed out in Weinberg's book  \cite{Weinberg:2008zzc}\footnote{At page 476.}. Ignoring that  issue,  the  account  concludes by indicating that these perturbations continue evolving into the cosmic structure responsible for galaxy formation, stars, planets and eventually life and human beings.

One  of the  least  satisfactory features of the  inflationary scenario  is  that it seems to require  the  incorporation of new  fields  or new  degrees of freedom\footnote{Starobinsky inflationary scenario  seems,  at first  sight, to  bypass this  unattractive  feature,   however the point is that in higher derivatives theories ,  gravitation itself  has  more degrees of freedom than  standard Einstein's  gravity \cite{Myrzakulov:2014hca}. },  which  are not associated  with other existing manifestations beyond those  related to the inflationary process itself. It would  be certainly a much  more attractive scheme  one in which    inflation was tied  instead to one of the known  physical fields,  and thus  the   idea  that  the   role of the inflaton might be  played by the Higgs was  deemed  very attractive. However  the specific realization of such idea was faced with several problems related to the validity of the perturbation theory and the relative size of the radiative corrections \cite{Barbon:2009ya,Burgess:2010zq,Hertzberg:2010dc}.

On the other hand, and  connected in  a sense with Weinberg's  above mentioned   concern,   as it has been pointed out in various  articles \cite{Sudarsky:2013zoa,Canate:2012ua} and extensively discussed  \cite{Sudarsky:2009za}, there is a fundamental problem with the traditional point of view:  there  is  simply  no  clear  answer to the question   of  why,  how   and when did the   homogeneity and  isotropy of the  universe  break  down.  The  point is that both the   background, often described in a classical language,  and the perturbations  which  are described in a quantum language,  are both characterized  by homogeneous and isotropic states\footnote{That the quantum  vacuum state is  invariant  under translations  and rotations  can be  easily  demonstrated  by    applying to it   a ``spatial  displacement" or  a rotation operator   constructed for  the quantum field theory in question.  See  \cite{Perez:2005gh,Sudarsky:2009za,Castagnino:2014cpa}}.  Regarding the quantum aspect to  which the   above  mentioned fluctuations  refer to,  and     that presumably give rise to the   primordial inhomogeneities   and  anisotropies,    the   point is the following:  the unitary evolution (process \emph{U} \cite{Bell:1989qe}) of a quantum state  follows the  deterministic Schr\"odinger equation which  preserves    the symmetries of the initial  state  that    are also  symmetries of the action. By   contrast,   the process \emph{R} (reduction) \cite{Bell:1989qe}  associated  with  the  measurement of an observable, which  does force the system to  collapse to  one of the eigenstates of the measured observable in a indeterministic fashion, might break  those symmetries.  However, the \emph{R} process   can be called  upon only  when  a  measurement by an external  observer/apparatus   is  involved.   In other words, the problem  of  accounting for the   breakdown of  the symmetry  resides in how  we characterize the process that should be considered as a  \emph{measurement}. Without this characterization we cannot properly  explain the emergence of the  cosmological asymmetries when we have started with a completely homogeneous and isotropic state,  such as the Bunch-Davies vacuum (an related  states) in a FRW space-time, while  the dynamics don't break the symmetries.  This  is   evidently  a  manifestation of the so-called   measurement  problem in quantum theory, which in  fact becomes even more serious  when we want to apply the quantum theory to the universe as a whole,   simply because we can not even rely on the practical usage of identifying an observer and/or a measuring device. Simple  ``escape''  strategies,  like arguing that  we  as    astronomers  are acting as observers,  would not do resolve the issues as,  according to our understanding of   cosmology, we  ourselves  must be {\it consequence}  of the breakdown of symmetry that led to the generation of cosmological  structure and  thus cannot  also be its  {\it cause}. We will not  dwell  any further on  the discussion of these issues here, and after acknowledging that the  debate   is  not     completely  settled,   we   point the interested  reader to   the various   works  in which these issues  are examined in   detail and    where  the  references to dissenting  views  are cited \cite{Sudarsky:2009za,Okon:2013lsa}. 

The approach followed to address the issue, which was developed  in several previous  works \cite{DiezTejedor:2011ci}, is  based on the  incorporation into the inflationary paradigm of the hypothesis of self-induced collapse of the quantum  state, an idea employed in modified versions of quantum theory designed precisely to address the measurement problem \cite{Bell:1989qe,Bassi:2003gd}.  

In this  paper we  will show  that  the  incorporation of  the  spontaneous collapse  of quantum states into the theoretical framework   offers, as  a side  benefit, a   path  to   deal with  the  difficulties    faced  by the  Higgs-inflationary  scenario.      

The paper is organized  as  follows. In section I, we offer a   review  of  a  proposal   designed  to address  the  problem in the inflationary paradigm,  by providing a physical  account of how and when, starting  with homogeneous and isotropic state,  the   evolution  would  transform it into an state with  actual an-isotropic  and  in-homogeneous  perturbations  (rather  than  simple   quantum fluctuations  which  should  more  properly  be referred  to as quantum uncertainties). In section II,  we  provide a brief review of Higgs inflation to point out  some  of the problems faced by the theory. Finally in section III, we will incorporate the Continuous Spontaneous Localization formalism into the Higgs inflation theory and  see  how  the scenario described  in section II is modified. We end  with a brief  discussion  of our   results.

\section{Inflation with a spontaneous collapse theory}
The standard  approach is not able to transform quantum uncertainties into  the actual inhomogeneities and anisotropies  in standard  inflationary  cosmology. However, incorporating the most  promising   ideas  to deal  with the standard  measurement problem is able to do so. Therefore we consider a modification of the standard quantum mechanics that  incorporates a new dynamical feature responsible for the collapse of the wave function. As   has  been   shown in \cite{Sudarsky:2013zoa,Canate:2012ua} this approach can be applied to cosmology and  it offers  a  reasonable  account  for  the emergence of  the primordial  inhomogeneities and  the anisotropies that we  see  in the CMB.

The starting point is a standard  inflationary model  based on  the action of a scalar field   minimally coupled to gravity, \cite{Guth:1980zm,Starobinsky:1979ty} 
\begin{equation}
\label{accion1}
S = \int d^4x \sqrt{-g} \left[ \frac{M_P^2}{2} R  - \frac{1}{2} g^{ab} \nabla_a \phi \nabla_b \phi - V(\phi) \right] \ .
\end{equation}

The standard  analysis  proceeds  by separating the metric into a spatially homogeneous and isotropic Friedman-Robertson-Walker background   treated   classically and its perturbation (to be treated  quantum mechanically), 
\begin{equation}
g_{ab} = g_{ab}^{(0)} + \delta g_{ab} \ .
\end{equation}
The background Friedman-Robertson-Walker space-time is  described  by the metric
\begin{equation}
ds_0^2 = a^2(\eta) \left[ -d\eta^2 + \delta_{ij} dx^i dx^j \right] \ , \quad d\eta = dt / a \ .
\end{equation}
Using  the conformal Newton gauge and ignoring the vector and tensor part of the metric perturbations, one can write the  perturbed  metric  as 
\begin{equation}
ds^2 = a(\eta)^2 \left[-(1+2\Phi) d\eta^2 + (1-2\Psi) \delta_{ij} dx^i dx^j \right] \ ,
\end{equation}
where $\Phi$ and $\Psi$ are functions of the space-time coordinates $\eta, x^i$. 
A  similar  separation
\begin{equation}
\phi(t,\vec{x}) = \phi_0(t) + \delta \phi(t,\vec{x}) \ . 
\end{equation}
is considered for the scalar field $\phi$. Also, the scale factor is  approximated by  $a(\eta) \simeq -1/(\mathcal{H} \eta(1-\epsilon))$ with $\mathcal{H} \equiv a'/a = aH$  where  $H$  is the Hubble parameter, which during inflation is approximately constant. 

The main deviation from a perfect de Sitter expansion is given by the \emph{slow-roll} parameter $\epsilon$ defined as $$\epsilon \equiv \frac{M_P^2}{2} \left(\frac{\partial_\phi V}{V}\right)^2 \ .$$ During inflation, the energy density of the Universe is  taken to be  dominated by the inflation potential $V(\phi)$, and during  the  regime of slow-roll inflation the \emph{slow-roll} parameter could be written as
\begin{equation}
\label{epsilonH}
\epsilon = 1 - \frac{\mathcal{H}'}{\mathcal{H}^2} \ll 1 \ ,
\end{equation}

At this point the standard procedure is to consider  the Mukanov-Sasaki variable $$v \equiv a \left(\delta \phi + \frac{\phi_0'}{\mathcal{H}} \Psi \right) \ , $$ and quantize it in order to obtain  the power spectra of the metric from $\langle 0 | \hat{\Psi}(\eta,\mathbf{x}) \hat{\Psi}(\eta,\mathbf{y}) | 0 \rangle$, which, as we noted later, is identified  with the  ensemble  average $\overline{\Psi(\eta,\mathbf{x}) \Psi(\eta,\mathbf{y})}$ over a set  of  classical configurations. However this last step is   precisely  the one   that is  not  really justified,  as  was  noted  by  Weinberg \cite{Weinberg:2008zzc}.

We believe that the correct way to compute the power spectra is to calculate the  estimate  for   the  value of  ${\Psi}(\eta,\mathbf{x})$\footnote{ This  is  in fact the quantity that  appears in      the  semi-classical  Einstein's   equations as  seen in  equation (\ref{perturcslk})  below. } and then take the average of the product $\overline{{\Psi}(\eta,\mathbf{x}) {\Psi}(\eta,\mathbf{x}') }$ over the ensemble of possible universes.  The point is that  the    actual    analysis  of the  observed   spectra is  extracted  from    data of the actual temperature  departure from  isotropy   as given  by  the  relative temperature    deviation  $\delta T (\theta, \varphi) /T$ from the mean  at   each  specific  angle.   We  should be  able to characterize  a single universe, at least in principle,  and to   clarify   what  such characterization means,  before we    go on to   characterize  the ensemble\footnote{The discussion of   the conceptual pitfalls    often     encountered in attempts to    address  the issue  withing the standard  inflationary approach,  which  often rely on unjustified   calls to ``specific  realizations of stochastic   variables " that  however play no role in   the standard quantum theory  unless a measurement is  involved,    while  at the same  time  ignoring the  difficulties   that would represent   calling for a measurement carried out  by human  hands in the present context, are presented   in detail  in \cite{Sudarsky:2009za, Castagnino:2014cpa}.}.   

However, such   procedure  is  simply  not possible in the standard approach,  simply  because it  would  yield
\begin{eqnarray}
\langle 0 |  \hat{\Psi}(\eta,\mathbf{x})|0 \rangle = 0 \ .
\end{eqnarray} 
Since the vacuum state is homogeneous and isotropic and since the dynamics preserves the symmetries of the theory we have
that   same result for all  times. This result is problematic as we want to identify  somehow  our estimates   for
$ \Psi$ with the seeds of the inhomogeneities in the CMB. There are analogous instances where quantum theory presents us with a symmetric quantum state whereas nature exhibits asymmetric behavior. Perhaps the best known example is the   decay of a $J = 0$ nucleus.  See  a discussion  of that  example  in \cite{Castagnino:2014cpa}. Although the wave function is rotationally invariant, the alpha particles are seen to move on linear trajectories. The problem was studied by Mott \cite{Mott79}, and was resolved by heavy usage of the collapse postulate appropriate to a measurement situation \cite{Sudarsky:2009za}. However, in the cosmological problem at hand, even if we wanted to, we could not achieve a similar explanation, for we cannot call upon any external entity making a measurement. Therefore a new approach is needed.

We start in a semi-classical regime of gravity where the metric is coupled to the expectation value of the scalar field's energy-momentum tensor according to $$ G_{ab} = 8 \pi G \langle \hat{T}_{ab} \rangle \ .$$ So, by considering the equation of motion to first order in the perturbations, $\delta G_{ab} = 8 \pi G  \delta T_{ab} $, the equation for the Newtonian potential, in the slow-roll regime $(\epsilon \simeq 0)$, is given by
\begin{equation}
\label{perturcslk}
\nabla^2 \Psi(\eta,\mathbf{x}) =  \frac{4 \pi G \ \phi_0'(\eta)}{a} \langle \hat{\pi}(\eta,\mathbf{x}) \rangle \ ,
\end{equation}
where $ \hat{\pi}(\eta,\mathbf{x}) =  a \delta \hat{\phi}'(\eta,\mathbf{x})$ is the conjugate momenta.

\subsection{Continuous Spontaneous Localization}
\label{csltheory}

CSL theory was first proposed by Philip Pearle \cite{Pearle:2012tf}. This theory describes the collapse of a quantum state towards an eigenvector of the operator $\hat{A}$ with rate $\lambda_c$ \cite{Pearle:2012tf} and is characterized  in terms of two main equations. The first equation is a modified Schr\"odinger equation whose solution is
\begin{equation}
\label{cslfintvec}
|\phi,t\rangle_w = \mathrm{T} \exp \left( - \int_0^t dt' \left\{i \hat{H} + \frac{1}{4 \lambda_c} \left[w(t') - 2 \lambda_c \hat{A}\right]^2 \right\} \right) |\phi,0 \rangle \ ,
\end{equation}
where $\mathrm{T} $ is the ``time  ordering'' operator,  $\hat A$   is a self adjoint operator (usually called  the collapse operator) which  determines the preferential  basis to which the collapse  dynamics drives an arbitrary initial state, and $w(t)$ a white noise random function responsible for the  stochasticity  in the eventual evolution of the quantum system in question. The  second equation is the joint probability of the independent increments in $dw(t)$ at successive values of $t$, given by
\begin{equation}
\label{cslfintprob}
P(w) Dw = {}_w\langle \phi,t | \phi,t \rangle_w \prod_{t'=0}^{t-dt} \frac{dw(t')}{\sqrt{2\pi \lambda_c / dt}} \ .
\end{equation}
We have then that for every white function $w(t)$ there is a state vector given by $|\phi,t\rangle_w$ (\ref{cslfintvec}) which happens at Nature with the probability given by equation (\ref{cslfintprob}). The state vector evolves following this scheme and, according to (\ref{cslfintprob}), the vectors with largest norm are  the most probable since the evolution is non-unitary.  Such dynamics give us an ensemble of different evolutions for the state vector, where each one is characterized by a function $w(t)$. 

In standard quantum mechanics, the unitary condition is mandatory in order to guarantee that the sum of probabilities is equal to unity. However, in CSL there is no necessity to impose such condition, but only that in this scheme the total probability over all the possible $w(t)$'s be unity.

\subsection{CSL as a mechanism to generate quantum perturbations}
\label{dinamicanocolapso}
Various works \cite{Martin:2012pea, Canate:2012ua, DiezTejedor:2011ci,Das:2013qwa} have offered a treatment of the inflationary account for the emergence of the seeds of cosmic structure based on adaptations of the CSL theory. We  will follow  the one presented in \cite{Canate:2012ua}, whose  approach is a semi-classical treatment, where quantum fields are treated  quantum  mechanically   while space-time  degrees of freedom  are  described in a classical  language. That  combination  has sometimes been considered as unviable \cite{Page:1981aj}, whose  basic  argument is that   semi-classical gravity  without collapse of the wave function   leads to  conflicts  with the  experiment  they  considered,  while  a collapse where  the  expectation value of  the energy  momentum tensor jumps  would be  in conflict with the   semi-classical Einstein  equation. In previous  works \cite{Canate:2012ua,Sudarsky:2013zoa} we  have   advocated   an approach in which semi-classical Einstein's  equation is  viewed  as  an approximate description of   limited  validity  in analogy  with,  say, the   Navier-Stokes description  of a fluid  (for a more  detailed   discussion  see \cite{Sudarsky:2013zoa}) .  This  led  to  the development of a self-consistent formalism   allowing the   phenomenological incorporation of spontaneous collapse  \cite{DiezTejedor:2011ci} in  which  the semi-classical treatment has  been  shown  to  be  viable for  certain applications,  including in  particular  the  inflationary context  at hand. 

 Here we offer a  short  review  of the specific treatment used in \cite{ Canate:2012ua},  of the  standard   (i.e.  inflaton distinct from the Higgs) inflationary  situation    which will    serve as a basis for our analysis of Higgs  inflation. The treatment  considers  the perturbation of the   inflaton as  a quantum field  in a classical space-time, characterized initially by the adiabatic  vacuum  state,  with the  collapse leading to  a change in   quantum state, that in turn leads to the   emergence of the  primordial  inhomogeneities  and  anisotropies  of the metric. 

The quantity that is measured is  $\Delta T / \overline{T}$, which is a function of the coordinates ($\theta, \varphi$) on the celestial two-sphere. This data is expanded in terms of spherical harmonics as
\begin{equation}
\label{DTharmonic}
\frac{\Delta T}{\overline{T}}(\theta,\varphi) = \sum_{lm} \alpha_{lm} Y_{lm} \ ,
\end{equation}
where the expansion coefficients $\alpha_{lm}$ are given by
\begin{equation}
\alpha_{lm} = \int d^2\Omega \ \frac{\Delta T}{\overline{T}} Y_{lm}^* \ .
\end{equation}
Considering the first order perturbations in the Fourier basis for  (\ref{perturcslk})
\begin{equation}
\Psi(\eta,\mathbf{k}) = - \frac{4 \pi G \chi_0'(\eta)}{a k^2} \langle \hat{\pi} (\eta,\mathbf{k}) \rangle \ ,
\end{equation}
and the well known Sachs effect together with other local effects concerning the  emission of photons by the cooling plasma \cite{Mukhanov:2005sc} leads  to the well known relation \cite{Mukhanov:1990me}
$$ \frac{\Delta T}{\overline{T}} \simeq \frac{1}{3} \Psi(\eta, x).$$
Thus  one  can express the coefficients $\alpha_{lm}$, which are the quantities of direct experimental interest for the observations, as
\begin{equation}
\alpha_{lm} = c\int d^2 \Omega  \ Y_{lm}^* \int d^3k \ \frac{e^{i \mathbf{k} \cdot \mathbf{x}} }{k^2} \langle \hat{\pi}(\mathbf{k},\eta) \rangle \ , 
\end{equation}
\begin{equation}
c \equiv - \frac{4 \pi G \ \phi_0'(\eta)}{3 a}  \ . 
\end{equation}
We will be  evaluating  the relevant  quantities at the end of inflation, $\eta = -\tau$ (see Appendix \ref{estimates}), but this  ignores several  effects related to the post reheating period  including  amplification  and  plasma  oscillations  that need to  be  treated  by suitable   adjustments at the end of our calculations \cite{Leon:2010fi}. The measures that are relevant for the quantification of cosmological fluctuations are the averages of the expansion coefficients $\alpha_{lm}$ over the ensembles of possible universes,  corresponding in  our case to the   possible realizations of the stochastic  functions  appearing in the CSL  dynamics. These averages can be expressed as
\begin{equation}
\overline{|\alpha_{lm}|^2} = (4\pi c)^2 \int_0^{\infty} \frac{dk}{k} \ j_l^2(k R_D) \frac{\overline{\langle \hat{\pi}(\mathbf{k},\eta) \rangle^2}}{k} \ ,
\end{equation}
where 
\begin{equation}
 \overline{\langle \hat{\pi}(\mathbf{k},\eta) \rangle \langle \hat{\pi}(\mathbf{k}',\eta) \rangle^*} \equiv \overline{\langle \hat{\pi} \rangle^2} \delta(\mathbf{k}-\mathbf{k}') \ .
\end{equation}

As we have mentioned, in the standard treatment $ \overline{\langle \hat{\pi} \rangle^2} = 0$ (not to be confused with $\overline{\langle \hat{\pi}^2 \rangle}$). However, with the CSL formalism it has been shown \cite{ Canate:2012ua} that with suitable choices of the collapse generating operators  (corresponding to the    objects $ \hat{A}$  appearing in equation (\ref{cslfintvec}))  one obtains $\overline{\langle \hat{\pi} \rangle^2}=\alpha k$, with  $\alpha$  a constant, which translates into a scale invariant spectrum
\begin{equation}
\overline{|\alpha_{lm}|^2}  = \alpha (4\pi c)^2 \frac{1}{2l(l+1)} \ .
\end{equation}

In fact,  as it is shown in   detain in  \cite{Canate:2012ua}, using the CSL formalism, the prediction is given by
\begin{equation}
\label{paverensem}
\overline{\langle \hat{\pi} \rangle^2} \simeq \frac{\tilde{\lambda}_c k \mathcal{T}}{2}  \ ,
\end{equation}
when the collapse operator appearing in (\ref{cslfintvec}) is $\hat{A} = \hat{\pi} = a \delta \hat{\phi}'$. Similar results are obtained if one considers $\hat{A} = \hat{x} = a \delta \hat{\phi}$. The parameter $\mathcal{T}$ is the time at which inflation starts and is computed in the Appendix \ref{estimates}. 

Thus, by computing $\overline{\langle \hat{\pi}(\mathbf{k},\eta) \rangle^2}$ in the framework of CSL theory, we are able to compute values of the $\alpha_{lm}$ and then, by using equation (\ref{DTharmonic}), predict the mean square temperature fluctuations at a point in the sky to be
\begin{equation}
\label{csltemppert}
\overline{\left( \frac{\Delta T}{T} \right)^2} = \frac{\pi}{4} \left[ \frac{4\pi G \phi'_0}{3 a} \right]^2 \tilde{\lambda}_c \mathcal{T} \mathcal{I} \ ,
\end{equation}
where $\mathcal{I}$ characterizes  the  range of  co-moving wave numbers that are relevant in the observed CMB,
\begin{equation}
\mathcal{I} \equiv \frac{1}{\tilde{\lambda}_c} \int dk \lambda_c(k) \leq \int_{10^{-3}}^{10^2} \frac{dk}{k} \simeq 10 \ .
\end{equation}

The slow-roll parameter given by equation (\ref{epsilonH}) can be rewritten in terms of the field and its potential as
\begin{equation}
\epsilon \simeq \frac{3 \phi_0'^2}{a^2 V(\phi_0)} \ ,
\end{equation}
therefore, the term
\begin{equation}
\left[ \frac{4\pi G \phi'_0}{3 a} \right]^2 \simeq \frac{\epsilon V}{M_P^4} \ ,
\end{equation}
and we can rewrite the expression for the temperature fluctuations (\ref{csltemppert}) as
\begin{equation}
\label{cslepsilon}
\overline{\left( \frac{\Delta T}{\bar{T}} \right)^2} = \frac{\epsilon V}{M_P^4} \tilde{\lambda}_c \mathcal{T} \mathcal{I} .
\end{equation}
However, in order to compare  with observations   we  need to   consider  the effects of the  post  reheating  epoch  in  the   estimate (\ref{cslepsilon}). As shown in \cite{Leon:2013pya}, the main effect on the overall amplitude of the   fluctuations  is  to multiply de previous result by $1/\epsilon^2$. Thus, we have finally
\begin{equation}
\overline{\left(\frac{\Delta T}{\bar{T}}\right)^2} \simeq \frac{V}{\epsilon M_P^4} \tilde{\lambda}_c \mathcal{T} \mathcal{I}  \ .
\end{equation}

We can use cosmological data to put constraints on the parameters of the CSL model.  The quantity $\left(\Delta T/\bar{T}\right)^2$ is determined  by the observations to be of order  $10^{-10}$ \cite{Planck:2013jfk}, so by considering a small value for the slow roll parameter $\epsilon \simeq 10^{-2}$ and considering $V^{1/4} \simeq 10^{15} \text{GeV} \simeq 10^{-4} M_P$ given by the GUT scale,  we have $$\tilde{\lambda}_c \mathcal{T} \sim 10^3 \ . $$ 

As  shown in Appendix \ref{estimates}, the conformal time $\mathcal{T}$ is  of order $10^{17}$ s, and thus  we  must  have
\begin{equation}
\tilde{\lambda}_c \sim 10^{-14} \ s^{-1} \ .
\end{equation}
This value is not far away from the $10^{-16} \ s^{-1}$ suggested by GRW \cite{Ghirardi:1987ns}. These   results   indicate  that   the  incorporation of  the  CSL  modifications to  quantum  dynamics leads  to a  suitable characterization of the   generation of the  primordial inhomogeneities  with  predictions that  can match the observations.

\section{Higgs Inflation}
\label{higgsinflation}
In recent years, there have been efforts  to develop  schemes  in which  the Higgs field of the pure Standard Model  could play the  role of the inflaton field. One of the main obstacles faced  by that program  has to do with the following. Models where the inflaton field is minimally coupled to gravity can produce the observed density perturbations only  if  the scalar field has   a mass of  about $\sim 10^{13}$ GeV or a very small coupling constant $\lambda \sim 10^{-13}$ \cite{Planck:2013jfk}. However  the Higgs field of  the Standard  Model has a coupling constant of the order $\lambda = m_h^2/v^2 \simeq \mathcal{O}(1) $  and a mass of 125 GeV.  This problem can be overcome through  the introduction of a non-minimal coupling such that the usual constraints over the coupling constant $\lambda$ might be  relaxed. Considering a non-minimal coupling  allows  inflation to last long enough by making the Higgs  effective  potential  very  flat and thus  produce the adequate number  of e-folds \cite{Fakir:1990hm,Bezrukov:2008bp}. Here  we   review   the essence  of  these treatments  which  will  serve as a basis  upon  which   our modified  analysis  will be built.
 
The action considered for gravity  and  the Higgs field  with non-minimally coupling, in the Jordan frame, is 
\begin{equation}
\label{Sphi}
S_J[h] = \int d^4 x  \sqrt{-g} \bigg( \frac{M_P^2 + \xi h^2}{2} R  - \frac{\nabla^a h \nabla_a h}{2} - \frac{\lambda}{4}(h^2-v^2)^2 \bigg) + \mathcal{L}(\Psi_{matter}, g)\ , 
\end{equation}
where $M_P = (8 \pi G)^{-1/2}$ is the Planck mass, $h$ is the Higgs scalar field with  ¨vacuum expectation value¨ $v$ and $\xi$ is the non-minimal coupling constant. According to the work  by Bezrukov \cite{Bezrukov:2007ep}, the analysis of the previous action is better understood by passing to the Einstein frame, through the change of  variables 
\begin{equation}
\tilde{g}_{\mu \nu} = \Omega^2  g_{\mu \nu}, \quad \Omega^2 = 1 + \frac{\xi h^2}{M_p^2} \ ,
\end{equation}
where $\Omega$ is known as the conformal factor. The action in the Einstein frame is then
\begin{equation}
S_E [\chi] = \int d^4 x \sqrt{- \tilde{g}} \left\{ \frac{M_P^2}{2}\tilde{R} - \partial_\mu \chi \partial^\mu \chi - U(\chi) \right\} \ ,
\end{equation}
where $\tilde{R}$ is the Ricci  scalar  of the  metric $\tilde{g}_{\mu \nu}$. Here  we have  also  introduced the new field $\chi$,  defined by
\begin{equation}
\label{xihvar}
\frac{d \chi}{d h} = \sqrt{\frac{\Omega^2 + 6 \xi^2 h^2 / M_P^2}{\Omega^4}} \ ,
\end{equation}
 and   the  new potential  term 
\begin{equation}
\label{Upot}
U(\chi) = \frac{\lambda}{4 \Omega^4(h)} (h^2(\chi) - v^2)^2 \ .
\end{equation}
 The equation  (\ref{xihvar}) has an exact solution \cite{Bezrukov:2013fka}, although little information can be read from it. However, in   particular regimes  for the value of the field $h$, the differential equation (\ref{xihvar}) has  simple explicit solutions for $\chi(h)$, given by
\begin{equation}
\label{inflationscale}
\chi \simeq
\begin{cases}
h & \quad \text{if } h \ll \frac{M_P}{\sqrt{\xi}} \ ,\\
\sqrt{\frac{3}{2}} M_P \ln \Omega^2(h) & \quad \text{if } h \gg \frac{M_P}{\sqrt{\xi}} \ .\\
\end{cases}
\end{equation}
Considering the second condition, $h \gg M_P/\sqrt{\xi}$, we have that the form for the new Higgs potential is
\begin{equation}
\label{uplano}
U(\chi) = \frac{\lambda M_P^4}{4 \xi^2} \left( 1 - \text{exp} \left[ \frac{- 2 \chi}{\sqrt{6} M_P} \right] \right)^{2} \ ,
\end{equation}
which is flat when $\chi \gg M_P$, making chaotic inflation possible in the slow-roll regime.

The conformal transformation allows the recovery  of  the usual formalism for inflation where the slow-roll parameters are now defined by
\begin{equation}
\label{epplano}
\epsilon = \frac{M_P^2}{2} \left( \frac{dU /d\chi}{U} \right) ^2 \simeq \frac{4 M_P^4}{3 \xi^2 h^4} \ ,
\end{equation}
\begin{equation}
\label{etaplano}
\eta = M_P^2 \frac{d^2U /d\chi^2}{U} \simeq - \frac{4 M_P^4}{3 \xi^2 h^4} \left(1-\frac{\xi h^2}{M_P^2}\right) \ .
\end{equation}
In the usual scheme, inflation ends when $\epsilon \simeq 1$, so that, according to (\ref{epplano}), the value of the field at the end of inflation must be 
\begin{equation}
\label{hend}
h_{end} = \left(\frac{4}{3}\right)^{1/4} \frac{M_P}{\sqrt{\xi}} \simeq 1.07 \frac{M_P}{\sqrt{\xi}} \ ,
\end{equation}
or $\chi_{end} \simeq 0.94 \ M_P$.
Also, the number $\tilde{N}$ of \emph{e-folds}, in the Einstein frame, is determined by
\begin{eqnarray}
\label{efoldxi}
\tilde{N} & = & \int_{h_{end}}^{h_N} \frac{1}{M_P^2} \frac{dU}{dU/dh} \left(\frac{d \chi}{d h}\right)^2 dh \nonumber \\ 
& \simeq & \frac{3}{4} \left[ \left(\xi + \frac{1}{6}\right) \frac{h_N^2 - h_{end}^2}{M_P^2} - \ln \left( \frac{M_P^2+\xi h_N^2}{M_P^2+\xi h_{end}^2}\right) \right] \ , 
\end{eqnarray}
where $h_N$ is determined by the number of inflationary e-foldings in the Einstein frame. In the context of Higgs inflation,  in the regime  where $\xi h^2 \gg M_P^2$, the last expression is simplified to
\begin{equation}
\label{efoldxiaprox}
\tilde{N} \simeq \frac{3}{4} \frac{h_N^2 - h_{end}^2}{M_P^2/\xi} \ .
\end{equation}
Then, the value of $h_N$ is determined by the number of \emph{e-folds}  required to have the right amount of inflation. In this case one  usually  considers 60 \emph{e-folds}, leading to
\begin{equation}
\label{hNvalue}
h_N \simeq 9.14 \frac{M_P}{\sqrt{\xi}} \ .
\end{equation}

Now that we have translated from the Jordan to the Einstein frame according to (\ref{uplano}), (\ref{efoldxiaprox}) and obtained the value of the field  at the   end  of  inflation (\ref{hNvalue}), one  can estimate  the  amplitude  of the   scalar   metric  perturbations  as  in the standard treatments. The result is
\begin{equation}
\label{powermetric}
\mathcal{P}_\Psi (k,\eta) = \frac{U(h_N)}{24 \pi^2 M_P^4 \epsilon} \ .
\end{equation}
Experimentally, we have  $\mathcal{P}_\Psi \simeq 10^{-10}$, therefore one  would   require
\begin{equation}
\label{espotforHI}
\frac{U}{\epsilon} \simeq (0.0276 \ M_P)^4 \ .
\end{equation}
By substituting the expression for $U(h_N)$ given by  (\ref{uplano}), one  finds  that  
\begin{equation}
\label{nonmincons}
\frac{\lambda}{\xi^2} \simeq 10^{-10} ,
\end{equation}
and for $\lambda \simeq 1$ the  required value of the non-minimal coupling constant is
\begin{equation}
\label{xiconst}
| \xi | \simeq 10^4 - 10^5 \ .
\end{equation}
The resulting model  is  known  to lead to a successful inflation scenario producing the spectrum of primordial fluctuations in agreement with the observational data. However, the self consistency of this model has been questioned in several papers \cite{Burgess:2009ea,Hertzberg:2010dc,Lerner:2009na}. 

It is natural to wonder if a coupling as large as $\xi \simeq 10^4$ could jeopardize the validity of the classical approximation on which the inflation behavior is based. As it is shown in \cite{Burgess:2009ea,Atkins:2010eq}, standard power-counting techniques imply that semi-classical perturbation theory must break down at energies of the scale $$\Lambda = \frac{M_P}{\xi} \ ,$$ so this scale can be regarded as an upper bound on the energy over which the theory can be considered an effective field theory.

The point is  that, from equation (\ref{inflationscale}), one can read the energy scale at which Higgs  inflation takes place,  and find it to be   given  by 
\begin{equation}
\Lambda_{\text{HI}} = \frac{M_P}{\sqrt{\xi}} \ .
\end{equation}
As $\xi \gg 1$, then $\Lambda_{\text{HI}}>\Lambda$ and this would imply that  the  theory could not be trusted in the energy scale at which  inflation  is   supposed to  take place.

There seems  to be a straightforward  path  for  solving this problem  by forcing  the value of $\xi$ to be less than  unity in order to invert the inequality $\Lambda_{\text{HI}} > \Lambda$. However, in standard Higgs inflation theory this is problematic  because,   as   seen in equation (\ref{xiconst}),  $\xi$ is constrained by the observations. As we will see the situation is drastically modified when collapse theories are relied on to generate the actual inhomogeneities and anisotropies.

\section{Higgs Inflation with CSL}
\label{higgscsl}

As we discussed in the previous  section, the Higgs Inflation model with non-minimal coupling can, through a conformal transformation, be cast in a form that resembles  the standard treatment of the inflaton coupled minimally to gravity. Here  we proceed  to implement the incorporation  of  the CSL mechanism in that model   treating the problem in the  Einstein frame, so that  starting point is  the action for the scalar field $\chi$ coupled to gravity, written as 
\begin{equation}
\label{einsteinaction}
S_E [\chi] = \int d^4 x \sqrt{- \tilde{g}} \left\{ \frac{M_P^2}{2}\tilde{R} - \frac{1}{2} \tilde{g}^{ab} \partial_a \chi \partial_b \chi - U(\chi) \right\} \ .
\end{equation}
Following the usual procedure, one obtains   the semi-classical equation resulting from (\ref{perturcslk}) for the first order perturbations in the Fourier basis, resulting in
\begin{equation}
-  k^2\Psi(\eta,\mathbf{k}) = \frac{4 \pi G \chi_0'(\eta)}{a } \langle \hat{\pi} (\eta,\mathbf{k}) \rangle \ .
\end{equation}
Taking the same steps as presented in \cite{Canate:2012ua}, one obtains the  prediction  for  the CMB fluctuations in the temperature, which now take the form
\begin{equation}
\overline{\left(\frac{\Delta T}{\bar{T}}\right)^2_{\chi_0} } \simeq \frac{U(\chi_0)}{\epsilon M_P^4} \tilde{\lambda}_c \tilde{\mathcal{T}} \mathcal{I} \sim 10^{-10}  \ , \quad  U(\chi_0) = \frac{\lambda M_P^4}{4 \xi^2} \ .
\end{equation}	
 Considering  the orders of magnitude $\epsilon \simeq 10^{-2}$ and $\mathcal{I} \sim 10$, we estimate for the combination
\begin{equation}
\frac{\tilde{\mathcal{T}} \tilde{\lambda}_c}{4 \xi^2} \simeq 10^{-13} \ ,
\end{equation}
and, according to (\ref{conftimeini}), we obtain the relation for $\tilde{N}$, the number of e-folds in the Einstein frame,
\begin{eqnarray}
\frac{\tilde{\lambda}_c}{\xi} & \simeq & 3 \ e^{-\tilde{N}} \times 10^{2} \ \text{s}^{-1} \ .
\end{eqnarray}
Then, in order to  obtain a model where  $\xi < 1$, the rate of collapse must satisfy 
\begin{equation}
\label{lambdabound}
\tilde{\lambda}_c \leq \ e^{-\tilde{N}} \times 10^{2} \ \text{s}^{-1} \ . 
\end{equation}
One last aspect to consider is that $\tilde{N}$ is related to the number $N$ of e-folds in the Jordan frame \cite{Bezrukov:2013fca} according to 
\begin{equation}
\label{NtotildeN}
\tilde{N} = \ln \left(\frac{\tilde{a}_{end}}{\tilde{a}}\right) = N + \ln \left( \frac{1}{\tilde{N}} \right) \ ,
\end{equation}
where $\tilde{a}_{end}$ is the scale factor when inflation ends given in the Einstein frame. 

\begin{figure}[h]
	\begin{center}
		\includegraphics[width=0.8\linewidth]{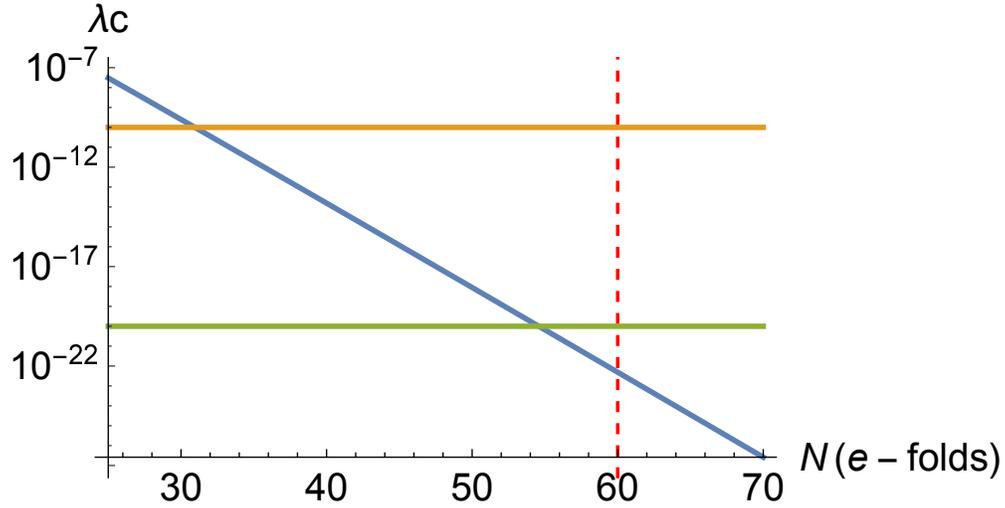}
		\caption{Relation between the physical e-folds $N$ (Jordan frame) and the collapse parameter $\lambda_c(s^{-1})$. We can appreciate the regions where $\lambda_c$ in permitted according to \cite{Carlesso:2016khv}. Also, the central value of 60 e-folds correspond to a value of $\lambda_c(s^{-1}) \simeq 10^{-22}$.}
		\label{lambdaefold}
	\end{center}
\end{figure}

Thus, considering the previous equation and  focusing on   the equality in (\ref{lambdabound}), we obtain  a direct relation between the number of  physical e-folds and the collapse parameter $\tilde{\lambda}_c$ which is presented in Figure \ref{lambdaefold}. Also the upper and lower bounds on the  standard   CSL collapse parameter are indicated, which according to \cite{Carlesso:2016khv} characterizes   the experimental constraints  as  extracted  directly from  laboratory  tests   involving   non-relativistic   many   particle  systems.  These  constraints place  the collapse parameter in the range   $10^{-20}$ to $10^{-10}$ that according to  our  analysis  correspond to a range of 30 -- 55 e-folds.

Also, according to Figure \ref{lambdaefold}, for  inflation to last for $N\simeq 60$ e-folds in the physical frame, the collapse rate   would  need to be $\tilde{\lambda}_c \simeq 10^{-22} \ \text{s}^{-1} $ .

\section{Discussion}
The inflationary paradigm has been quite successful in estimating  the   shape  and   scale of  the   primordial  spectrum of fluctuations, which ultimately  are seen   in temperature patterns  in the CMB,  with some of its  dominant  features also  identified   in the large  scale  distribution of  galaxies. However, as we have emphasized throughout this work, the standard treatment does not really account for the emergence of the primordial inhomogeneities. The new approach that has been proposed subjects the quantum  dynamics of the  field associated  with  the  fluctuations of the inflaton to a collapse mechanism, specifically described  through  an  adaptation to cosmology of the CSL theory.

By considering this new approach for treating the quantum perturbations of the inflation field, we are able to relax the conditions that strongly constrain Higgs inflation,  maintaining the analysis   in the regime where perturbation theory  can  still  be  considered  valid. 

We have been able to solve the problem in a natural  way  while   allowing the non-minimal coupling constant $\xi$   introduced  in these contexts to be less than 1. This has been achieved while    considering  a   range  of values  of  the collapse rate $\lambda_c$  which, according to Figure \ref{lambdaefold}, are   related   directly to  the  maximum number $N$ of e-folds  expected in the corresponding situation (and which  are  compatible  with $\xi  <1$ ). The range  that is   considered  as viable,  in view  of experimental  constrains and  the  suitability of the theories in resolving the measurement problem, for $\lambda_c$ results in  a range of 30 -- 55 e-folds,   which   is   very  close  to the   commonly assumed range of 50 -- 60 e-folds, as considered  for instance  in equation 24 of \cite{Planck:2013jfk}.

The fact that we  have considered the collapse dynamics as parametrized  in terms of  the   cosmological time  associated with the Einstein frame  can  be regarded  as  indicating  that  the collapse parameter  would  depend on ``time"  when  the evolution is expressed   in  terms of the cosmological time  associated with the Jordan  frame. This, in turn, might be seen  as tied to the proposals  made  in recent papers  suggesting that  the collapse rate   might  depend  on   curvature \cite{Modak:2016uwr, Okon:2013lsa}. Such feature would be relevant for cosmological applications because,  during the inflationary regime, the  Universe's  curvature  would have  been  quite different from that prevailing in the laboratory conditions to  which  the  relevant  bounds  on $\lambda_c$  do apply. Thus,  the  inclusion  of  the novel idea  in the  analysis  could  be  justified  and   be  relevant  in   the   establishment   of the viability for Higgs-Inflation and the  adjustment  of the non-minimal coupling constant $\xi$ in the context of  these  proposals.

In summary, we have considered a simple  version of  the  incorporation of  the  self-induced collapse of the wave function as a mechanism to generate, during the inflationary epoch the actual  primordial inhomogeneities  and  anisotropies  which  are supposed  to seed  the  structures in the Universe. By considering the  incorporation of  collapse theories  into the  picture,  we were able to reassess the prospects  that the Higgs field might be the field driving inflation, allowing  to bypass the  problems that afflict the  standard accounts of  Higgs inflation and which  constitute  the main reason to  consider  it unviable. As a byproduct, we found  an interesting relationship between  the parameter $\lambda_c$ characterizing the (fixed) collapse  rate of the CSL collapse theory  with the maximum number of e-folds allowed by the model.

\section{Acknowledgments}

This research project was supported  CONACyT  fellowship  during S. R.'s PhD program.

D.S. is supported in part by the CONACYT grant No. 101712. and by UNAM-PAPIIT grant IN107412. 

\appendix

\section{Estimates of physical quantities}
\label{estimates}

In order to relate  the value of the collapse rate $\lambda_c$  to the expected   value  for the  amplitude of the   power spectrum  of primordial inhomogeneities,   we need to estimate the values of the conformal time $\eta$ at which inflation ends, $- \tau$, and starts, $- \mathcal{T}$.  Recalling that the temperature of radiation scales like $1/a$, and assuming that the effective temperature at the end of inflation corresponds to the GUT scale given by $10^{15}$ GeV and the radiation temperature today is $2.7 \ \text{K} = 2.4 \times 10^{13}$ GeV, we can estimate the scale factor $a(-\tau)$ according to
\begin{eqnarray}
\label{scaletau}
\frac{T(\text{today})}{T(-\tau)} = \frac{2.4 \times 10^{-13} \ \text{GeV}}{10^{15} \ \text{GeV}} = \frac{a(-\tau)}{1}  \ , 
\end{eqnarray}
yielding
\begin{equation}
 a(-\tau) = 2.4 \times 10^{-28} \ .
\end{equation}

In order to find the value of $\eta=-\tau$ that corresponds to the previous scale factor we can use the Friedmann equation $$H^2 = \frac{8\pi G}{3} V(\phi_0) \ , \quad V^{1/4} \sim \text{GUT scale} $$ to determine
\begin{equation}
H \simeq \frac{M^2_{GUT}}{3 M_P} \simeq 10^{13} \ \text{GeV} \ ,
\end{equation}
where $M_P = (8 \pi G)^{-1/2} = 2.435 \times 10^{18} \ \text{GeV}$ and $M_{GUT} \simeq 10^{16}$.

Solving the classical equations of motion, the scale factor corresponding to the inflationary era is, to a good approximation,
\begin{equation}
a(\eta) = \frac{-1}{H \eta} \ ,
\end{equation}
and so we can evaluate $a(\tau)$ to obtain  $\tau = 10^{-24} Mpc$ .

We can also compute the value of the conformal time at the start of inflation, for this we can assume that inflation lasts 60 e-folds, thus
\begin{equation}
\frac{\mathcal{T}}{\tau} = \frac{a(-\tau)}{a(-\mathcal{T})} = e^{60} \simeq 10^{26} \ ,
\end{equation}
therefore $\mathcal{T} = 10^2$ Mpc.

\subsection{Estimates in the Einstein frame}
The same estimations are needed in the section \ref{higgscsl}, but in the Einstein frame, since we have performed a conformal transformation in order to have the scalar inflaton field minimally coupled to gravity. The transformation from the physical frame (Jordan frame) to the Einstein frame is given by
\begin{eqnarray}
d \tilde{s}^2 & = & \tilde{a}^2(\tilde{\eta}) [ -d\tilde{\eta}^2 + d\tilde{\vec{x}}^2 ] \ , \nonumber \\
& = & \Omega^2 a^2(\eta) [-d\eta^2 + d\vec{x}^2] \ ,
\end{eqnarray}
where the conformal factor $\Omega$ is given by
\begin{equation}
\Omega^2(h) = 1 +\frac{\xi h^2}{M_P^2} \ .
\end{equation}

First we need to estimate the value of $\Omega$ at the end of inflation. So, we need the value of the field $h_{end}$ that we have already computed in equation (\ref{hend}),
\begin{equation}
h_{end} = \left(\frac{4}{3}\right)^{1/4} \frac{M_P}{\sqrt{\xi}} \ .
\end{equation}
Therefore, the value of $\Omega$ at the end of inflation $(\eta=-\tau)$ is
\begin{equation}
\label{omegaend}
\Omega(-\tau) \simeq  \frac{3}{2} \ .
\end{equation}
The next step is to use the Friedmann equation for the action given by (\ref{einsteinaction}), which is given by
\begin{equation}
\tilde{H}^2 = \frac{8\pi G}{3} U(\chi) \ .
\end{equation}
The value of the potential $U(\chi)$ is given by (\ref{uplano}) and can be evaluated at $\chi_{end} = 9.14 M_P$, resulting in
\begin{equation}
\quad U(\chi_{end}) \simeq \frac{\lambda M_P^4}{16 \xi^2} \ ,
\end{equation}
and, substituting into the previous equation, we have
\begin{equation}
\tilde{H}^2 = \frac{1}{48} \frac{M_P^2}{\xi^2} \quad \Rightarrow \quad \tilde{H} \simeq \frac{7}{2 \xi} \times 10^{17} \text{GeV} \ .
\end{equation}
Now, we use the solution for the scale factor in this frame, given by
\begin{equation}
\tilde{a}(\tilde{\eta}) = \frac{-1}{\tilde{H} \tilde{\eta}} \ , \quad \tilde{a}(\tilde{\eta}) = \Omega(\eta) a(\eta) \ ,
\end{equation}
evaluating at the time when inflation ends, $\tilde{\eta} = -\tilde{\tau}$, we can solve for $\tilde{\tau}$, yielding
\begin{equation}
\tilde{\tau} = 8 \xi \times 10^9 \ \text{GeV}^{-1} \ ,
\end{equation}
where we have used the value of $a(-\tau)$ given by (\ref{scaletau}).
With this results we can compute the value of the conformal time at the start of inflation assuming that inflation lasts N e-folds, thus we have
\begin{equation}
\label{conftimeini}
\frac{\tilde{\mathcal{T}}}{\tilde{\tau}} = \frac{\tilde{a}(-\tilde{\tau})}{\tilde{a}(-\tilde{\mathcal{T}})} = e^{\tilde{N}} \quad \Rightarrow \quad \tilde{\mathcal{T}} = 8 \xi e^{\tilde{N}} \times 10^9 \ \text{GeV}^{-1} \ .
\end{equation} 
This approximation of $\mathcal{T}$ is used in Section \ref{higgscsl} to compute the fluctuation in the temperature.

\bibliography{referenceshiggs}

\end{document}